\documentclass{article}
\usepackage{emulateapj}
\usepackage{apjfonts}
\usepackage{graphics}

\newenvironment{inlinefigure}{%
\def\@captype{figure}%
\noindent\begin{minipage}{0.999\linewidth}\begin{center}}
{\end{center}\end{minipage}\smallskip}

\newlength{\colwidth}
\newcommand{\Mpc}{\hbox{{\rm Mpc}}}

\newcommand{\hydra}{{\sc{HYDRA }}}

\newcommand{\lya}{{\mbox Ly$\alpha$~}}

\renewcommand{\H}{\ion{H}{1}}

\newcommand{\He}{\ion{He}{1}}
\newcommand{\Hep}{\ion{He}{2}}
\newcommand{\Hepp}{\ion{He}{3}}
\newcommand{\h}{{\rm H\,{\sc I}}}

\newcommand{\hep}{{\rm He\,{\sc II}}}
\newcommand{\hepp}{{\rm He\,{\sc III}}}

\newcommand{\mtau}{{\hbox{$\bar\tau_{\rm eff}$}}}
\newcommand{\tauGP}{{\hbox{$\tau_{\rm GP}$}}}
\newcommand{\be}{\begin{equation}}
\newcommand{\ee}{\end{equation}}
\newcommand{\yr}{\hbox{{\rm yr}}}

\newcommand{\kms}{\hbox{{\rm km}\,{\rm s}$^{-1}$}}

\lefthead{Theuns et al.}
\righthead{Detection of \Hep\ reionization in the SDSS}

\begin{document}


\title{Detection of \Hep\ reionization in the SDSS quasar sample}
\author{Tom Theuns\altaffilmark{1,2}, Mariangela
Bernardi\altaffilmark{3,4}, Joshua Frieman\altaffilmark{4,5}, Paul
Hewett\altaffilmark{1}, Joop Schaye\altaffilmark{6}, Ravi
K. Sheth\altaffilmark{7,5} and Mark Subburao\altaffilmark{4}}

\altaffiltext{1} {Institute of Astronomy, Madingley Road, Cambridge CB3
0HA, UK}
\altaffiltext{2} {Universitaire Instelling Antwerpen, Universiteitsplein
1, B-2610 Antwerpen, Belgium}
\altaffiltext{3} {Department of Physics, Carnegie Mellon University, 5000 Forbes Ave.
Pittsburgh, PA 15213}
\altaffiltext{4} {University of Chicago, Astronomy and Astrophysics
Center, 5640 S. Ellis Ave., Chicago, IL 60637}
\altaffiltext{5} {Fermi National Accelerator Laboratory, P.O. Box 500,
Batavia, IL 60510}
\altaffiltext{6} {School of Natural Sciences, Institute for Advanced Study, Einstein Drive, Princeton NJ 08540}
\altaffiltext{7} {Department of Physics and Astronomy, University of
Pittsburgh, Pittsburgh, PA 15620}

\begin{abstract}
The expansion of the Universe leads to a rapid drop in the hydrogen
\lya\ effective optical depth of the intergalactic medium (IGM),
$\mtau\propto (1+z)^{3.8}$, between redshifts 4 and 3. Measurements of
the temperature evolution of the IGM and of the \Hep\ opacity both
suggest that \Hep\ reionizes in this redshift range. We use
hydrodynamical simulations to show that the observed temperature
increase associated with \Hep\ reionization leads to a relatively
sudden decrease in $\mtau$ around the reionization epoch of $\approx$
10 per cent. We find clear evidence for such a feature in the evolution
of $\mtau$ determined from a sample of $\sim 1100$ quasars obtained
from the SDSS. \Hep\ reionization starts at redshift $\approx 3.4$, and
lasts for $\Delta z \approx 0.4$. The increase in the IGM temperature
also explains the widths of hydrogen absorption lines as measured in
high-resolution spectra.
\end{abstract}


\section{Introduction}
Neutral hydrogen in the intergalactic medium (IGM) resonantly scatters
the flux blueward of the hydrogen \lya emission line in quasar spectra
(Gunn \& Peterson 1965; Bahcall \& Salpeter 1965; Peebles 1993, \S
23). The fact that not all the flux is absorbed implies that the IGM is
very highly ionized. The neutral fraction is determined from the
balance between photo-ionizations, produced by the UV-background
radiation from galaxies and quasars, and recombinations. For hydrogen,
the recombination rate depends on temperature $\propto 1/T^{0.7}$, and
so an increase in temperature will lead to a decrease in the level of
absorption. Recent evidence suggests that \Hep\ reionizes around a
redshift $z\sim$3--3.5, and the associated temperature increase should
have a measurable effect on the mean absorption. In this letter, we
compute the optical depth evolution of a simulation in which \Hep\
reionization heats the IGM at redshift 3.4. The resulting dip in the
evolution of the mean absorption matches that recently measured by
Bernardi et al.\ (2002) in the Sloan Digital Sky Survey (SDSS) data.

Hydrodynamical simulations have proved to be very successful in
reproducing many properties of the observed absorption (Cen et
al. 1994; Zhang, Anninos \& Norman 1995; Miralda-Escud\'e et al.\ 1996;
M\"ucket et al.\ 1996; Hernquist et al.\ 1996; Theuns et al.\ 1998;
Machacek et al.\ 2000; see e.g. Efstathiou, Schaye and Theuns (2000)
for a recent review). Most of the absorption at redshifts 2--5 arises
in modestly over- and under dense filamentary and sheetlike structures,
leading to a forest of \lya-absorption lines in the spectra of quasars
(Lynds 1971). These structures can be described and understood with
simple physical models in which the typical size of the absorbers is
determined by the Jeans length of the photo-heated gas (Bi \& Davidsen
1997; Schaye 2001). Given that these hydrodynamical simulations
reproduce the data in great detail, we are confident that they will
also allow us to investigate the effect of a sudden increase in $T$ on
the mean absorption.

Several lines of recent evidence suggest that \Hep\ reionizes around
$z\sim $ 3--3.5, significantly later than hydrogen. Observations of the
\Hep\ \lya-forest detect a sudden increase in the mean \Hep\ opacity
around $z\sim 3$ (Reimers et al.\ 1996; Heap et al.\ 2000; Kriss et
al. 2001; Smette et al.\ 2002). Such a rapid increase in the flux of
\Hep\ ionizing photons is expected to occur at the final stages of
reionization when individual \Hepp\ bubbles around sources percolate
(Gnedin 2000). The resulting hardening of the ionizing background could
be responsible for the observed jump in the relative abundance of C{\sc
IV}/Si{\sc IV} (Songaila \& Cowie 1996), although more recent data (Kim
et al.\ 2002) apparently do not seem to support such a change (see also
Dav\'e et al.\ 1998). Finally, entropy injection associated with \Hep\
reionization will increase the temperature of the IGM (Miralda-Escud\'e
\& Rees 1994), which has the effect of making the \lya-absorption lines
broader on average. The gas will become nearly isothermal when the
change in temperature is large. Schaye et al.\ (2000) detected both
signatures in high-resolution Keck data, using a method based on the
finding of Schaye et al.\ (1999; see also Ricotti et al.\ 2000 and
Bryan \& Machacek 2000) that the variation of the widths of the
absorption lines as a function of column density, can be used to
measure the temperature as a function of the density. Theuns et al.\
(2002a) performed a wavelet analysis of the spectra of several high
resolution QSO spectra, and found that the data required a large
increase in $T$ by a factor $\sim 2$ around $z\sim 3.3$, consistent
with late \Hep\ reionization (Miralda-Escud\'e \& Rees 1994). The
temperature decrease after $z\sim 3.3$ is also consistent with this
interpretation (Theuns et al.\ 2002b). Other heating mechanisms have
been discussed in the literature as well, for example galactic winds
(Cen \& Bryan 2001). Although these may have contributed to the entropy
of the high redshift gas, it seems unlikely they could result in a
sudden change of temperature as seems required by the data.

In the next section, we use hydrodynamical simulations to investigate
the effect of \Hep\ reionization on the evolution of the mean effective
optical depth, \mtau\ . Section 3 summarizes the method used by Bernardi
et al.\ (2002) to measure $\mtau(z)$ from a sample of 1061 SDSS quasar
spectra, and section 4 compares the data to the simulations.

\section{Hydrodynamical simulations}
We will use hydrodynamical simulations to compute the change in the
effective optical depth \mtau, i.e. the ratio between the observed and
emitted fluxes in the \lya-forest region of a quasar spectrum,
\be
\exp(-\mtau) \equiv \langle {F_{\rm observed}\over F_{\rm
emitted}}\rangle=\langle\exp(-\tau)\rangle\,,
\ee
as the temperature increases during \Hep\ reionization. The
Gunn-Peterson optical depth \tauGP\ of a uniform IGM at redshift $z$ is
(see Peebles 1993, \S 23)
\be
\tauGP(z) = {\sigma_0 c x n_{\rm H}\over H(z)}\,,
\label{eq:tauGP1}
\ee
where $\sigma_0=(3\pi\sigma_T/8)^{1/2} f\lambda_0$, with $\sigma_T$
the Thomson cross-section, and $f=0.416$ and $\lambda_0=1215.67$\AA\ the
oscillator strength and wavelength of the hydrogen \lya-transition. The
Hubble constant is $H(z)$, $x=n_\h/n_{\rm H}$ is the neutral fraction,
and $n_{\rm H}$ is the hydrogen number density. In photo-ionization
equilibrium,
\be
x\approx (1+x_\hep+2x_\hepp) {n_{\rm H}\alpha(T)\over\Gamma}\,,
\label{eq:x}
\ee
where $x_\hep=n_\hep/n_{\rm H}$ and $x_\hepp=n_\hepp/n_{\rm H}$ are the
abundances of singly and doubly ionized helium, $\alpha\approx 3.975\times
10^{-13} (T/10^4{\rm K})^{-0.7}{\rm cm}^3 {\rm s}^{-1}$ is the hydrogen
recombination coefficient, and $\Gamma$ is the photo-ionization
rate. We have assumed the gas to be highly ionized, $x\ll 1$. Inserting
numerical values, and assuming a helium abundance of $Y=0.24$ by mass,
\begin{eqnarray}
\tauGP &=& 0.76\,({\Omega_m\over 0.3})^{-1/2}({h\over 0.65})^{-1}({T\over 10^4{\rm K}})^{-0.7}({\Gamma\over 10^{-12}{\rm
s}^{-1}})^{-1}\nonumber\\
&\quad&\,({\Omega_b h^2\over 0.02})^2\,({1+z\over
4})^{4.5}\,,
\end{eqnarray}
where $\Omega_bh^2$ is the baryon density, $\Omega_m$ the matter
density, and $H_0=100h$\kms\Mpc$^{-1}$ is the current Hubble
constant. Since the IGM is not uniform, the optical depth will
fluctuate, and the resulting fluctuating Gunn-Peterson absorption
produces the \lya-forest. This is what is computed in the
simulations. Because of the temperature dependence of the optical
depth, we would naively expect \mtau\ to decrease by $\approx
1/g^{0.7}$ as the temperature increases by a factor $g$ during \Hep\
reionization. We will show that the decrease is in fact less than that.

We have performed a simulation of a vacuum energy dominated,
cosmologically flat cold dark matter model, with cosmological
parameters $(\Omega_m,\Omega_bh^2, h,
\sigma_8,Y)$=(0.3,0.019,0.65,0.9,0.24). The simulations are performed
using a modified version of \hydra (Couchman, Thomas \& Pearce 1995;
Theuns et al.\ 1998), in which the gas is photo-heated by an imposed
uniform UV-background which evolves with redshift and reionizes \H\ and
\He\ at $z\sim 6$ and \Hep\ at $z\sim 3.4$.  During reionization of
both \H\ and \Hep\, we artificially increase the photo-heating rates --
i.e. the mean energy associated with each photo-ionization -- to mimic
the effects of radiative transfer (Abel \& Haehnelt 1999). The
temperature at the mean density, $T(z)$, of the simulation fits the
data of Schaye et al.\ (2000) and Theuns et al.\ (2002a), in particular
there is a sudden increase in $T$ at $z\sim 3.4$ by a factor $g\approx
1.8$ as a result of \Hep\ reionization. Non-equilibrium rates for
photo-ionization, cooling and heating are computed using the fits in
Theuns et al.\ (1998). The particle masses for dark matter and gas are
$1.1\times 10^7$ and $2.0\times 10^6M_\odot$ respectively, and the
simulation box is 12$h^{-1}$ (co-moving) Mpc on a side. To illustrate
the effect of \Hep\ reionization on \mtau, we use a similar simulation
in which \Hep\ reionization starts slightly later, at $z\sim 3.2$.
Note that these simulations assume that the radiation field is
spatially uniform at all times, see Sokasian, Abel \& Hernquist (2002)
for a \Hep\ reionization simulation that includes some radiative
transfer effects.
\begin{inlinefigure}
\label{fig:reion}
\centerline{\resizebox{0.96\colwidth}{!}{\includegraphics{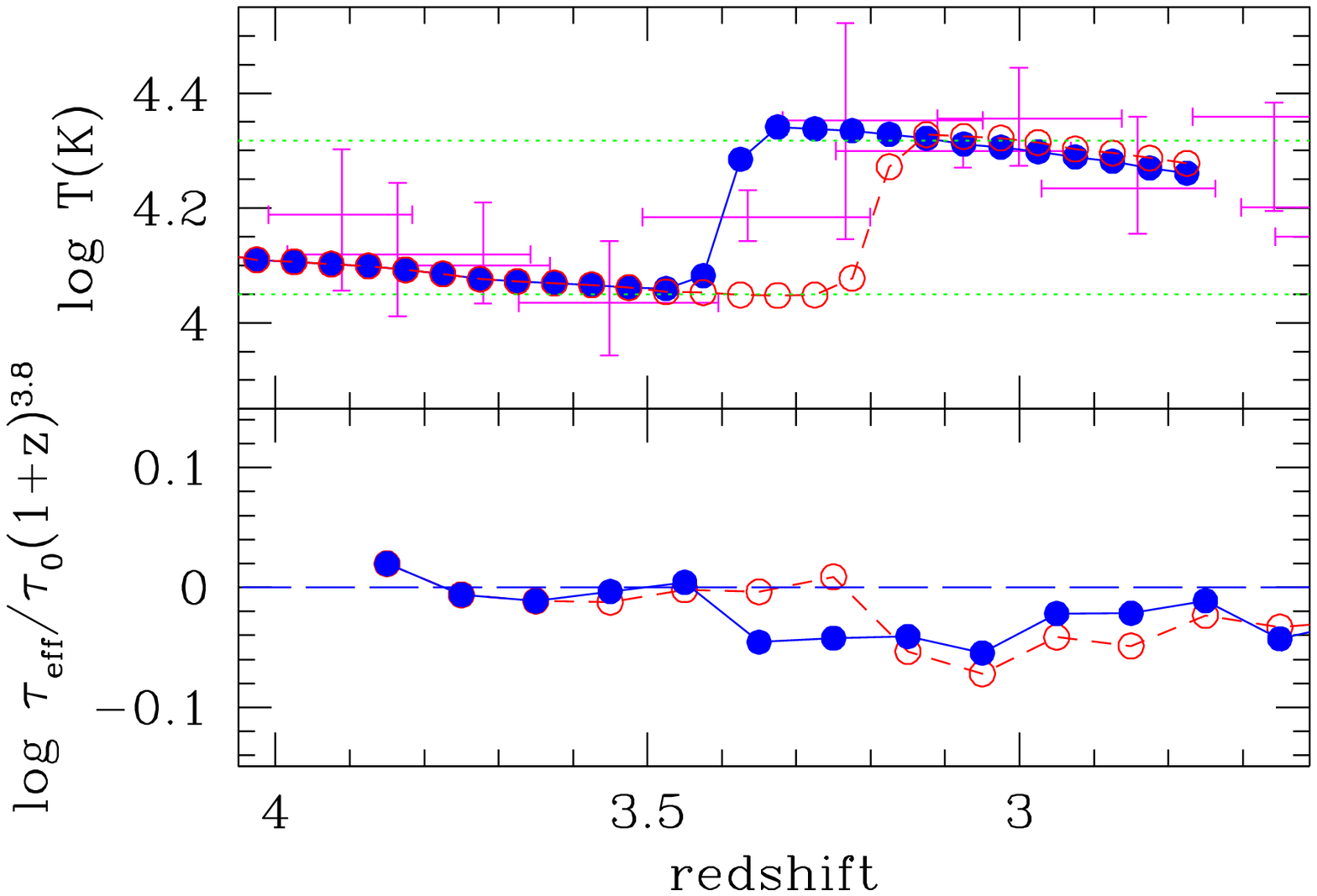}}}
\vspace{-1.5cm}
\figcaption{
Evolution of the temperature at the mean density $T$
(top panel), and the deviation of the effective optical depth \mtau\
from a power-law (bottom panel) for simulations in which \Hep\
reionization starts at $z\approx 3.4$ (full line and filled circles)
and $z\approx 3.2$ (dashed line and open circles). Data points with
error bars are the measurements from Schaye et al.\ (2000). \Hep\
reionization increases $T$ by a factor $\sim 1.8$ (dotted lines). Away
from reionization, $\mtau\propto (1+z)^\zeta$, with $\zeta\approx
3.8$. The increase in $T$ during reionization decreases the neutral
hydrogen fraction and hence also the mean absorption. This causes
\mtau\ to fall below the power law evolution, producing a feature in
$\mtau/(1+z)^\zeta$ that is characteristic for the occurrence of \Hep\
reionization.}
\end{inlinefigure}
While the simulation was running, we stored the data required to
compute mock spectra along random sight lines through the simulation
box, at many thousands of output times. The time interval between
individual snapshots is of order of the light crossing time through the
box. This fine sampling of the redshift evolution allows us to take
into account the detailed evolution of the density, temperature and
ionizing background. We impose the ionizing background $J_{\rm
{G+Q}}(\nu,z)$ from QSOs and galaxies as computed by Haardt \& Madau
(1996) and updated by Haardt \& Madau
(2002\footnote{http://pitto.mib.infn.it/\~{}haardt/cosmology.html}),
but reduced in amplitude by a factor of $j$. Using $j=2.75$ for all
redshifts $2.5\le 4$ reproduces the observed evolution of \mtau\ quite
well. This UV-background is different from the one used during the
simulation. However, because the hydrogen gas is in photoionisation
equilibrium, and because the hydrogen photo-heating rate is independent
of the hydrogen ionizing flux, such a rescaling works accurately (see
Theuns et al.\ 1998).

Using the mock spectra, we determine the evolution of \mtau, binned in
redshift intervals of order $\Delta z=0.1$, and use bootstrap
re-sampling to estimate the expected variance for the SDSS sample,
given the number of QSOs available. This scatter could be
underestimated due to missing large scale power in the simulation box,
but such a systematic effect is unlikely to affect the relative
decrement during and away from \Hep\ reionization.

In Fig.~1, we compare the evolution of $T$ and \mtau\ in the two
simulations, in which \Hep\ reionisation starts at redshifts 3.4 and
3.2, respectively. The temperature $T$ increases suddenly by a factor
$\sim 1.8$ following reionization. The effect of this temperature
increase on \mtau\ is shown in the bottom panel. Away from \Hep\
reionization, $\mtau\propto (1+z)^\zeta$, with $\zeta\approx 3.8$. The
bottom panel of the figure shows the deviation from this scaling, \mtau
$/(1+z)^\zeta$, and illustrates that the temperature increase leads to
a $\approx $ 10 percent decrease in \mtau. After $\Delta z\approx 0.4$,
\mtau\ recovers to the power-law evolution.

\begin{inlinefigure}
\centerline{\resizebox{0.96\colwidth}{!}{\includegraphics{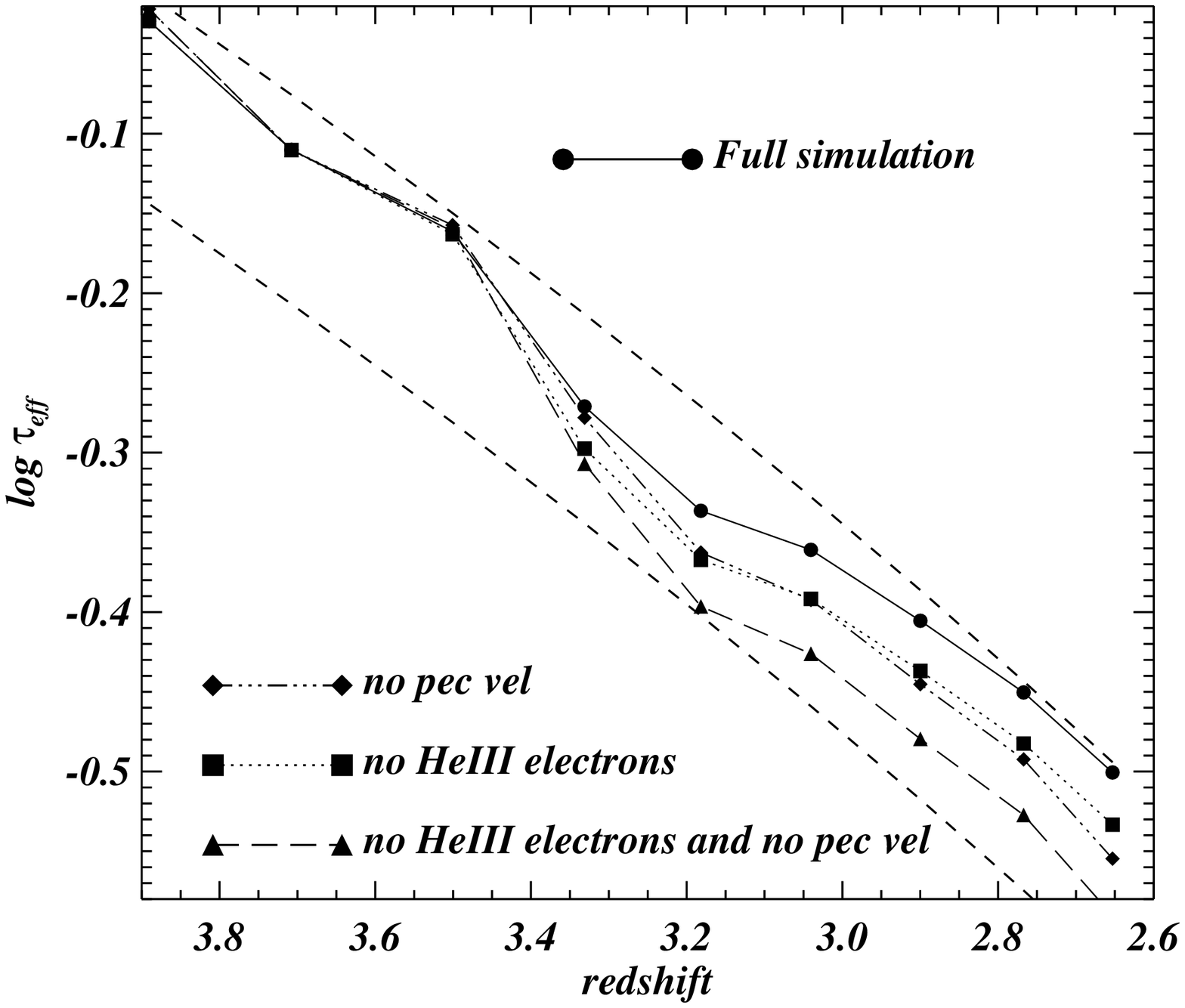}}}
\figcaption[fig:physics]{
Evolution of \mtau\ for the simulation with \Hep\
reionization at redshift 3.4 (circles).  The other curves illustrates
two reasons why the change in effective optical depth is smaller than
naively expected. If peculiar velocities are neglected in the
computation of the spectra (diamonds), then the mean absorption is
higher because gravitationally induced peculiar velocities tend to
increase blending, which in turn decreases the opacity. The curve has
been off-set vertically to match the full simulation curve above
$z=3.5$. During \Hep\ reionization, pressure induced peculiar
velocities push gas out of the filaments into the voids, and this
increases the absorption, and consequently the two curves start to
deviate. This shows that pressure induced peculiar velocities partly
counteract the decrease in \mtau\ . The increased electron density
following \Hep\ reionization, neglected in computing the dotted curve
(squares), has a similar effect on \mtau\ . When both peculiar
velocities and extra electrons are neglected (dashed curve connecting
triangles), then the decrease in \mtau\ is $\approx (1./1.54^{0.7})$
(parallel dashed lines), close to that expected from the temperature
dependence of the recombination coefficient.}
\label{fig:physics}
\end{inlinefigure}
This decrease is less than the $1/1.8^{0.7}\approx 30$ per cent naively
expected from the temperature dependence of the hydrogen recombination
coefficient. Note however, that \mtau\ does not scale linearly with the
optical depth, since $\exp(-\mtau)=\langle\exp(-\tau)\rangle$, and that
some of the absorbing gas is at densities higher than the mean, for
which the relative change in temperature is less (Schaye et al.\
2000). In addition, there are two other physical effects that lessen
the dependence of \mtau\ on $T$. The extra electrons liberated during
\Hep\ reionization tend to increase the neutral hydrogen abundance
(Eq.\ref{eq:x}). For full \Hep\ reionization, the {\em increase} in
\mtau\ is $\approx$ 7.5 per cent. Another effect is that the filaments
start to expand due to the extra heating (see Theuns, Schaye and
Haehnelt, 2000, Fig~(4)). These extra peculiar velocities also partly
counteract the decrease in \mtau\ .  Both effects are illustrated in
Fig.~2.

\begin{inlinefigure}
\centerline{\resizebox{0.96\colwidth}{!}{\includegraphics{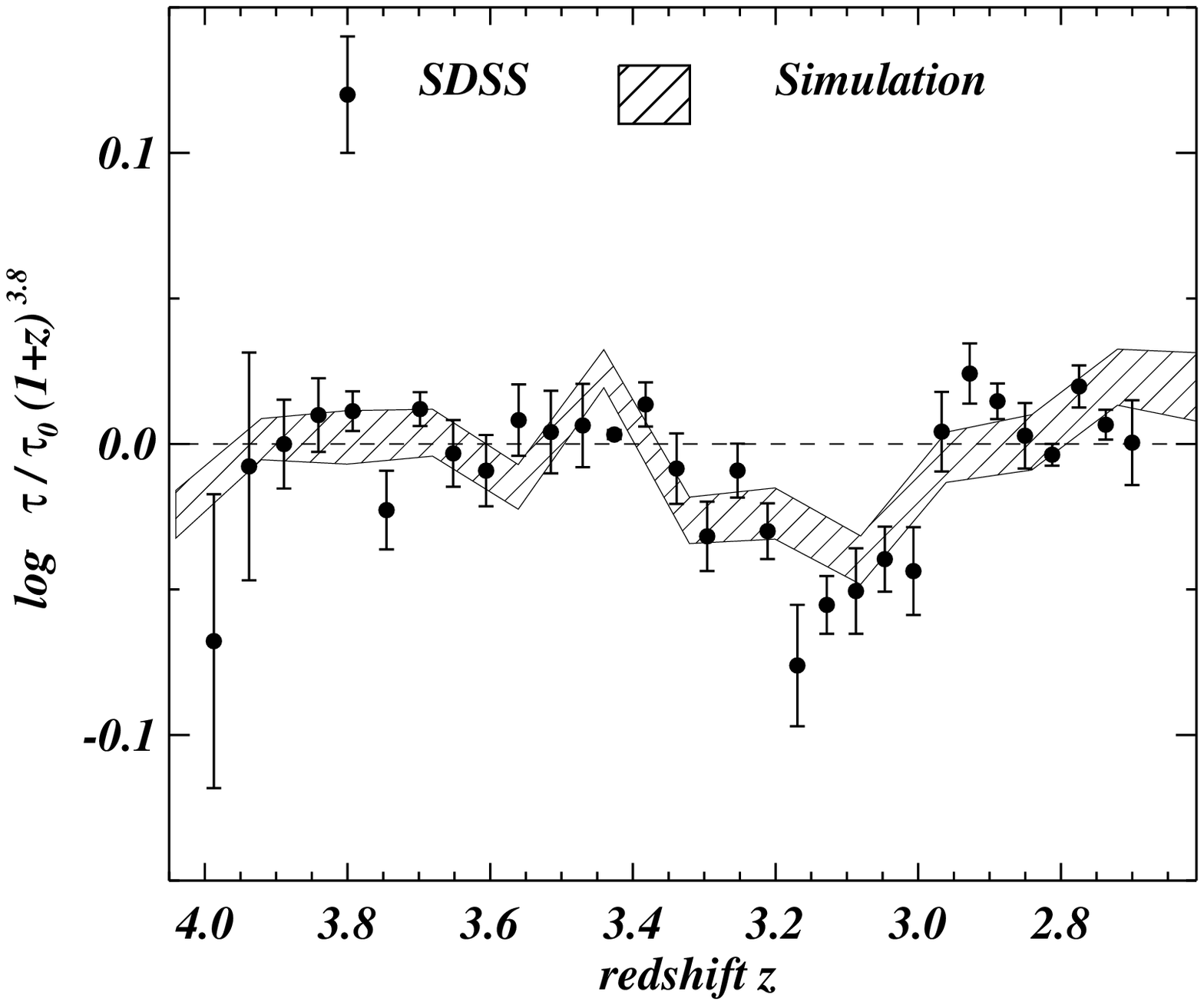}}}
\figcaption{
Deviation of the effective optical depth from a power-law evolution,
$\mtau/(1+z)^{3.8}$, for the SDSS data smoothed on 3000\kms (symbols
with error bars) and a hydrodynamical simulation of a $\Lambda$CDM
model, in which \Hep\ reionization starts at $z=3.4$ (hashed
region). The errors in the SDSS data are determined from bootstrap
re-sampling the individual spectra. The hashed region for the
simulation delineates the 20 and 80 per centile of bootstrap re-sampled
mock samples. The temperature increase associated with \Hep\
reionization causes \mtau\ to drop below the power-law evolution in the
simulation. This characterictic dip matches very well the feature
detected in the SDSS data.}
\label{fig:tau}
\end{inlinefigure}

\section{Determination of the \lya\ effective optical depth in the SDSS data}
Bernardi et al.\ (2002) determine \mtau(z)\ from 1061 moderate
resolution ($\lambda/\Delta\lambda\approx 1800$) and intermediate
signal-to-noise ratio (typically between 4 and 10) spectra of QSOs with
redshifts between 2.75 and 4.3, color selected from the SDSS
(Fig.~3). There are several crucial steps in the determination of the
mean absorption from these spectra, such as the determination of the
shape and amplitude of the continuum in each QSO spectrum, and
assessment of possible biases introduced by the QSO selection criterion
employed by the SDSS.

Full details of the procedure can be found in Bernardi et al
(2002). Briefly, the shape of the continuum in the \lya-forest region
is determined by solving simultaneously for the evolution of the
effective optical depth and for the shape of the mean continuum, while
allowing for the possibility that neither are well-fit by featureless
power-laws. This is possible, because the (intrinsic) shape of a
quasar's continuum is a function of {\em restframe} wavelength, whereas
\mtau\ is measured at given {\em observed} wavelength. Two different
methods were used, a $\chi^2$ technique, and an iterative fitting
procedure, and they give results in very good agreement with each
other. Both techniques allow one to take into account the presence of
weak emission lines in the QSO's continuum, and of deviations in the
evolution of the optical depth from a smooth power-law.

Bernardi et al.\ present many tests of their methods. For example, they
show that the scatter of individual measurements from the mean
evolution is uncorrelated with rest-frame wavelength, which is not the
case when the weak emission lines are not taken into account. The
evolution of \mtau\ determined from the SDSS data is consistent with
previous measurements, and the excellent statistics allow for the
detection of a localized feature in \mtau(z)\ of the order of 10 per
cent. The next section compares the SDSS data to the hydrodynamical
simulation.

\section{Results}
The deviation of \mtau\ from a power-law $\mtau/(1+z)^{3.8}$ is plotted
in Fig.~3 for the SDSS sample (symbols with error bars) and the
simulation for which \Hep\ reionization starts at $z\approx 3.4$
(hashed region). With respect to the power-law, the measured \mtau\
shows a distinctive dip around redshift $z=3.14\pm 0.03$ of depth
$\Delta\mtau=-0.09\pm 0.02$ and width $\Delta z=0.09\pm 0.02$ (see
Bernardi et al.\ 2002 for the determination of the errors), consistent
with the signature of \Hep\ reionization. The similarity between the
data and the simulation is striking. Note that the parameters for the
simulation were tuned to match the temperature evolution measured by
Schaye et al.\ (2000), and hence the dip in $\mtau$ is a prediction of
the model. The fact that the power-law evolution as measured from the
SDSS is the same as computed from the simulation, suggests that the
ionizing background $J_{\rm {G+Q}}(\nu,z)$ describes the evolution of
the photo-ionization rate well.

The amplitude, shape and duration of the deviation of
$\mtau/(1+z)^\zeta$ from a straight line is very similar for the SDSS
data and the hydrodynamical simulation. The onset of the decrease in
\mtau\ is not correctly modeled in the simulation, because the
UV-background is assumed to be uniform at all times and because the
rise in the UV-background does not take into account that photons are
consumed during the reionization process. This is partly counteracted
by the still relatively poor sampling $\Delta z\sim 0.1$ -- worse than
in the SDSS data -- of the simulation outputs, but better sampling
would show a much more rapid decrease in $\mtau$ in the simulations. In
reality, the temperature distribution will become inhomogeneous, as
some regions become ionized -- and hence hot -- before others do. The
spatial scale of these fluctuations is of the order of the size of a
new \Hepp\ region in a singly ionized and homogeneous \Hep\ Universe (
Miralda-Escud\'e, Haehnelt \& Rees 2000)
\begin{equation}
R = {16.8\over 1+z}\,\left( {\dot N_{\rm ph}\over 10^{56} {\rm s}^{-1}}
{t_{\rm QSO}\over 10^7\yr}({\Omega_b h^2\over
0.02})^{-1}\right)^{1/3}\Mpc\,,
\end{equation}
which corresponds to $\approx 1200\kms$ or $\Delta z=0.016$ at redshift
3 ($\dot N_{\rm ph}$ is the luminosity of the QSO in ionizing photons,
and $t_{\rm QSO}$ its age). Simulations that include radiative
transfer and can resolve the \lya-forest are required to model this
process more realistically.

In principle, it should be possible to detect the onset of \Hep\
reionization from the occurrence of such ionized, hot
\Hepp\ bubbles. (The \lq thermal proximity effect\rq.) Theuns et al's
(2002b) wavelet method did not detect such temperature fluctuations on
scales larger than 5000 \kms. However, they only attempted to identify
{\em individual} \Hepp\ regions, and did not quantify whether the
presence of such hotter regions can be detected
statistically. Zaldarriaga (2002) attempted to do just that, using the
spectrum of QSO 1422+231, but again no signal was detected. With more
data available now, such a detection may be possible.

In summary: the evolution of the effective optical depth \mtau,
measured in the SDSS quasar sample by Bernardi et al.\ (2002), deviates
suddenly by $\approx 10$ per cent at redshift $z\approx 3.1$ from a
smooth power-law. The mean absorption determined from a hydrodynamical
simulation of the IGM, in which \Hep\ reionization starts at redshift
3.4, provides an excellent match to this feature in \mtau. In this
simulation, the decrease in absorption is due to the temperature
dependence of the hydrogen recombination coefficient, and the required
temperature change due to \Hep\ reionization, is in good agreement with
the temperature measurements from Schaye et al.\ (2000) and Theuns et
al. (2002). We consider this strong evidence that \Hep\ reionization
has been detected in the SDSS data. The onset of reionization starts
around $z\sim 3.4$ and percolation, corresponding to the maximum
deviation of \mtau\ from a power-law, occurs at $z\sim 3.1$.

{\em Acknowledgments} TT thanks PPARC for the award of an Advanced
Fellowship. JS is supported by a grant from the W.M.Keck
Foundation. This research was conducted in cooperation with Silicon
Graphics/Cray Research utilizing the Origin 2000 super computer at the
Department of Applied Mathematics and Theoretical Physics in Cambridge.

{}


\begin{thebibliography}{}
\bibitem[Abel \& Haehnelt(1999)]{1999ApJ...520L..13A} Abel, T.~\& Haehnelt, 
M.~G.\ 1999, \apjl, 520, L13 

\bibitem[Bahcall \& Salpeter(1965)]{1965ApJ...142.1677B} Bahcall, J.~N.~\& 
Salpeter, E.~E.\ 1965, \apj, 142, 1677 

\bibitem[]{} Bernardi M, et al., 2002, submitted to AJ, preprint (astro-ph/0206293)

\bibitem[Bryan \& Machacek(2000)]{2000ApJ...534...57B} Bryan, G.~L.~\& 
Machacek, M.~E.\ 2000, \apj, 534, 57 

\bibitem[Bi \& Davidsen(1997)]{1997ApJ...479..523B} Bi, H.~\& Davidsen, 
A.~F.\ 1997, \apj, 479, 523 

\bibitem[Cen \& Bryan(2001)]{2001ApJ...546L..81C} Cen, R.~\& Bryan, G.~L.\ 
2001, \apjl, 546, L81 

\bibitem[]{} Cen, R., Miralda-Escud\'e, J., Ostriker, J.P., Rauch, M.,
1994, ApJ, 437, L9

\bibitem[Couchman, Thomas, \& Pearce(1995)]{1995ApJ...452..797C} Couchman, 
H.~M.~P., Thomas, P.~A., \& Pearce, F.~R.\ 1995, \apj, 452, 797 

\bibitem[Dav{\' e} et al.(1998)]{1998ApJ...509..661D} Dav{\' e}, R., 
Hellsten, U., Hernquist, L., Katz, N., \& Weinberg, D.~H.\ 1998, \apj, 509, 
661 

\bibitem[Efstathiou, Schaye, \& Theuns(2000)]{2000RSLPT.358.2049E} 
Efstathiou, G., Schaye, J., \& Theuns, T.\ 2000, Royal Society of London 
Philosophical Transactions Series, 358, 2049 

\bibitem[]{} Gnedin, N. Y., 2000, ApJ, 535, 530

\bibitem[Gunn \& Peterson(1965)]{1965ApJ...142.1633G} Gunn, J.~E.~\& 
Peterson, B.~A.\ 1965, \apj, 142, 1633 

\bibitem[Haardt \& Madau(1996)]{1996ApJ...461...20H} Haardt, F.~\& Madau, 
P.\ 1996, \apj, 461, 20 

\bibitem[Heap et al.(2000)]{2000ApJ...534...69H} Heap, S.~R., Williger, 
G.~M., Smette, A., Hubeny, I., Sahu, M.~S., Jenkins, E.~B., Tripp, T.~M., 
\& Winkler, J.~N.\ 2000, \apj, 534, 69 

\bibitem[]{} Hernquist, L., Katz, N.., Weinberg, D.H.,
Miralda-Escud\'e, J., 1996, ApJ, 457, L51

\bibitem[] {} Kim T-S, Cristiani S, D'Odorico S, 2002, A\& A, in press,
preprint (astro-ph/astro-ph)

\bibitem[Kriss et al.(2001)]{2001Sci...293.1112K} Kriss, G.~A.~et al.\ 
2001, Science, 293, 1112

\bibitem[Lynds(1971)]{1971ApJ...164L..73L} Lynds, R.\ 1971, \apjl, 164,
L73 

\bibitem[Machacek et al.(2000)]{2000ApJ...532..118M} Machacek, M.~E., 
Bryan, G.~L., Meiksin, A., Anninos, P., Thayer, D., Norman, M., \& Zhang, 
Y.\ 2000, \apj, 532, 118 

\bibitem[Miralda-Escude \& Rees(1994)]{1994MNRAS.266..343M} Miralda-Escude, 
J.~\& Rees, M.~J.\ 1994, \mnras, 266, 343 

\bibitem[]{} Miralda-Escud\'e, J., Cen, R., Ostriker, J.P., Rauch, M.,
1996, ApJ, 471, 582

\bibitem[Miralda-Escud{\' e}, Haehnelt, \& Rees(2000)]{2000ApJ...530....1M} 
Miralda-Escud{\' e}, J., Haehnelt, M., \& Rees, M.~J.\ 2000, \apj, 530, 1 

\bibitem[Muecket, Petitjean, Kates, \& Riediger(1996)]{1996A&A...308...17M} 
M\"ucket, J.~P., Petitjean, P., Kates, R.~E., \& Riediger, R.\ 1996, \aap, 
308, 17 

\bibitem[] {} Peebles PJE, 1993, Principles of Physical Cosmology,
Princeton University Press

\bibitem[]{} Reimers, D., Kohler, S., Wisotzki, L., Groote, D.,
Rodeiguez-Pascual, P., \& Warmsteker, W., 1997, A\&A, 327, 890.

\bibitem[Ricotti, Gnedin, \& Shull(2000)]{2000ApJ...534...41R} Ricotti, M., 
Gnedin, N.~Y., \& Shull, J.~M.\ 2000, \apj, 534, 41 

\bibitem[Schaye(2001)]{2001ApJ...559..507S} Schaye, J.\ 2001, \apj, 559, 
507 

\bibitem[Schaye, Theuns, Leonard, \& Efstathiou(1999)]{1999MNRAS.310...57S} 
Schaye, J., Theuns, T., Leonard, A., \& Efstathiou, G.\ 1999, \mnras, 310, 
57 

\bibitem[Schaye et al.(2000)]{2000MNRAS.318..817S} Schaye, J., Theuns, T., 
Rauch, M., Efstathiou, G., \& Sargent, W.~L.~W.\ 2000, \mnras, 318, 817 

\bibitem[Smette et al.(2002)]{2002ApJ...564..542S} Smette, A., Heap, S.~R., 
Williger, G.~M., Tripp, T.~M., Jenkins, E.~B., \& Songaila, A.\ 2002, \apj, 
564, 542 

\bibitem[Sokasian, Abel, \& Hernquist(2002)]{2002MNRAS.332..601S} Sokasian, 
A., Abel, T., \& Hernquist, L.\ 2002, \mnras, 332, 601 

\bibitem[Songaila \& Cowie(1996)]{1996AJ....112..335S} Songaila, A.~\& 
Cowie, L.~L.\ 1996, \aj, 112, 335 

\bibitem[Theuns et al.(1998)]{1998MNRAS.301..478T} Theuns, T., Leonard, A., 
Efstathiou, G., Pearce, F.~R., \& Thomas, P.~A.\ 1998, \mnras, 301, 478 

\bibitem[Theuns, Schaye, \& Haehnelt(2000)]{2000MNRAS.315..600T} Theuns, 
T., Schaye, J., \& Haehnelt, M.~G.\ 2000, \mnras, 315, 600 

\bibitem[Theuns et al.(2002)]{2002MNRAS.332..367T} Theuns, T., Zaroubi, S., 
Kim, T., Tzanavaris, P., \& Carswell, R.~F.\ 2002a, \mnras, 332, 367 

\bibitem[Theuns et al.(2002)]{2002ApJ...567L.103T} Theuns, T., Schaye, J., 
Zaroubi, S., Kim, T., Tzanavaris, P., \& Carswell, B.\ 2002b, \apjl, 567, 
L103 

\bibitem[Zaldarriaga(2002)]{2002ApJ...564..153Z} Zaldarriaga, M.\ 2002, 
\apj, 564, 153 

\bibitem[]{} Zhang, Y., Anninos, P., Norman, M.L., 1995, ApJ, 453, L57

\end{thebibliography}
\end{document}